\documentclass[pdflatex, sn-nature]{sn-jnl}

\usepackage{graphicx}%
\usepackage{multirow}%
\usepackage{amsmath,amssymb,amsfonts}%
\usepackage{amsthm}%
\usepackage{mathrsfs}%
\usepackage[title]{appendix}%
\usepackage{xcolor}%
\usepackage{textcomp}%
\usepackage{manyfoot}%
\usepackage{booktabs}%
\usepackage{algorithm}%
\usepackage{algorithmicx}%
\usepackage{algpseudocode}%
\usepackage{listings}%

\theoremstyle{thmstyleone}%
\theoremstyle{thmstyletwo}%

\theoremstyle{thmstylethree}%

\unnumbered

\begin{document}

\title[]{High-Precision Atmospheric Constraints for a Cool T Dwarf from JWST Spectroscopy}




\author*[1]{\fnm{Callie E.} \sur{Hood}}\email{cehood@ucsc.edu}

\author[1]{\fnm{Sagnick} \sur{Mukherjee}}

\author[1]{\fnm{Jonathan J.} \sur{Fortney}}

\author[2]{\fnm{Michael R.} \sur{Line}}

\author[3,14]{\fnm{Jacqueline K.} \sur{Faherty}}

\author[3,4]{\fnm{Sherelyn} \sur{Alejandro Merchan}}

\author[5]{\fnm{Ben} \sur{Burningham}}

\author[3]{\fnm{Genaro} \sur{Suárez}}

\author[6]{\fnm{Rocio} \sur{Kiman}}

\author[7]{\fnm{Jonathan} \sur{Gagn\'e}}

\author[8]{\fnm{Charles A.} \sur{Beichman}}

\author[3,9]{\fnm{Johanna M.} \sur{Vos}}

\author[3,10]{\fnm{Daniella} \sur{Bardalez Gagliuffi}}

\author[111]{\fnm{Aaron M.} \sur{Meisner}}

\author[12,13]{\fnm{Eileen C.} \sur{Gonzales}}

\affil*[1]{\orgdiv{Department of Astronomy and Astrophysics}, \orgname{University of California, Santa Cruz}, \orgaddress{\city{Santa Cruz}, \state{CA}, \postcode{95064}, \country{USA}}}

\affil[2]{\orgdiv{School of Earth and Space Exploration}, \orgname{Arizona State University}, \orgaddress{\city{Tempe}, \state{AZ}, \postcode{85287}, \country{USA}}}

\affil[3]{\orgdiv{Department of Astrophysics}, \orgname{American Museum of Natural History}, \orgaddress{\city{New York}, \state{NY}, \postcode{10024}, \country{USA}}}

\affil[4]{\orgdiv{Hunter College}, \orgname{City University of New York}, \orgaddress{\city{New York}, \state{NY}, \postcode{10065}, \country{USA}}}

\affil[5]{\orgdiv{Centre for Astrophysics Research, Department of Physics, Astronomy and Mathematics}, \orgname{University of Hertfordshire}, \orgaddress{\city{Hatfield}, \postcode{AL10 9AB}, \country{UK}}}

\affil[6]{\orgdiv{Department of Astronomy}, \orgname{California Institute of Technology}, \orgaddress{\city{Pasadena}, \state{CA}, \postcode{91125}, \country{USA}}}

\affil[7]{\orgdiv{Department}, \orgname{Plan\'etarium Rio Tinto Alcan}, \orgaddress{\street{Espace pour la Vie, 4801 av. Pierre-de Coubertin}, \city{Montr\'eal}, \country{Canada}}}

\affil[8]{\orgdiv{NASA Exoplanet Science Institute/IPAC, Jet Propulsion Laboratory}, \orgname{California Institute of Technology}, \orgaddress{\city{Pasadena}, \state{CA}, \postcode{91125}, \country{USA}}}

\affil[9]{\orgdiv{School of Physics, Trinity College Dublin}, \orgname{University of Dublin}, \orgaddress{\city{Dublin 2} \postcode{D02 XF86}, \country{Ireland}}}

\affil[10]{\orgdiv{Department of Physics and Astronomy}, \orgname{Amherst College}, \orgaddress{\city{Amherst}, \state{MA}, \postcode{01003}, \country{USA}}}

\affil[11]{\orgname{NSF's National Optical-Infrared Astronomy Research Laboratory}, \orgaddress{\city{Tuscon}, \state{AZ}, \postcode{85719}, \country{USA}}}

\affil[12]{\orgdiv{Department of Astronomy and Carl Sagan Institute}, \orgname{Cornell University}, \orgaddress{\city{Ithaca}, \state{NY}, \postcode{14853}, \country{USA}}}

\affil[13]{\orgdiv{Department of Physics and Astronomy}, \orgname{San Francisco State University}, \orgaddress{\city{San Francisco}, \state{CA}, \postcode{94132}, \country{USA}}}

\affil[14]{\orgdiv{Department of Physics}, \orgname{The Graduate Center City University of New York}, \orgaddress{\city{New York}, \postcode{10016}, \state{NY}, \country{USA}}}


\abstract{We present observations of the T8 dwarf 2MASS 0415-0935 with JWST’s NIRSpec spectrograph using the G395H grating ($\sim$ 2.87 - 5.14 $\mu$m). We perform the first atmospheric retrieval analysis at the maximum spectral resolution of NIRSpec (R$\sim$2700) and combine the spectrum with previous observations to study the 0.9-20 $\mu$m spectral energy distribution. We obtain precise constraints on chemical abundances ($\sim$0.02 dex) for a number of species  which complicate our understanding of disequilibrium chemistry, particularly for CO$_{2}$ and PH$_{3}$. Furthermore, we measure a $^{12}$CO/$^{13}$CO ratio of $\sim 97^{+9}_{-8}$, making 2MASS 0415-0935 the coldest ($\sim 760$ K) substellar object outside of our solar system  with a measured $^{12}$CO/$^{13}$CO ratio. This work shows promise for similar observations with JWST to provide precise abundances of major chemical species as well as isotopologues, allowing for new tests of our understanding of the formation and atmospheres of substellar objects.}




\maketitle

\section{Introduction}\label{sec:intro}
Brown dwarfs, intermediate in mass between gas giants and low-mass stars, act as critical tests of our knowledge of substellar atmospheres. Since brown dwarfs and directly imaged exoplanets span a similar range in effective temperature \cite{Faherty2016}, understanding the emergent spectra of brown dwarfs can inform our interpretation of planetary spectra. Comparison of observed brown dwarf spectra to radiative-convective equilibrium (RCE) models allow us to assess our understanding of the dominant processes that sculpt these atmospheres. For example, stronger or weaker features than expected in chemical equilibrium have been observed for molecules such as CO and NH$_{3}$, respectively, for a number of brown dwarfs \cite[e.g.,][]{Saumon2007,Miles2020}, demonstrating the need to include vertical mixing in models to accurately describe these atmospheres.

Data-driven Bayesian inverse or “atmospheric retrieval” methods have enabled abundance measurements for a number of important molecules like H$_{2}$O, CH$_{4}$, CO, NH$_{3}$ \cite[e.g.,][]{Line2017,Burningham2017}. Metallicities and elemental ratios like C/O can be calculated from these measured abundances, which when measured for a range of objects can be used to inform our understanding of formation pathways of planets and brown dwarfs \cite{Molliere2022}. Isotopologue ratios can provide additional leverage for distinguishing amongst potential formation scenarios of exoplanets and brown dwarfs, particularly with the James Webb Space Telescope (JWST) and upcoming instruments on extremely large telescopes (ELTs) \cite{Molliere2019,Morley2019}. For example, the ratio of $^{12}$CO/$^{13}$CO has been measured for three extrasolar objects, indicating enhanced $^{13}$CO relative to solar for planetary bodies compared to brown dwarfs, potentially from accretion of ices with enhanced $^{13}$CO from beyond the CO snow line \cite{Zhang2021,Zhang2021b,Line2021}.  Expanding the sample of objects with measured $^{12}$CO/$^{13}$CO ratios could enable a greater understanding of planetary versus brown dwarf formation pathways and timelines. 

However, significant challenges have arisen in fitting models to observed brown dwarf spectra. Fundamental parameters from fitting RCE grid models to observations can vary strongly with observed wavelength range or amongst differing model families \cite[e.g.,][]{Tannock2022,Lueber2023}.  Similarly, grid model fitting and atmospheric retrievals can yield very different parameter measurements when applied to the same set of observations \cite{ZJ2021,Zalesky2022}. Furthermore, disequilibrium chemistry models required to reproduce observed features from carbon- and nitrogen-bearing molecules in T and Y dwarfs predict strong PH$_{3}$ features, but only a potential detection has been reported \cite{Burgasser2023} despite several efforts \cite{Morley2018,Miles2020,Beiler2023}, suggesting a need for revisions to our understanding of disequilibrium and phosphorous chemistry. 

The effort to solve these discrepancies could benefit from observations with higher spectral resolution and signal-to-noise ratio, particularly beyond 3 $\mu$m where the strongest features from CO, CO$_{2}$, PH$_{3}$, and NH$_{3}$ are expected. As such, observations with JWST \cite{Rigby2023} represent the next frontier for atmospheric studies. Recently published observations of brown dwarfs with JWST’s Near Infrared Spectrograph (NIRSpec,\cite{Jakobsen2022}) and Mid Infrared Instrument (MIRI, \cite{Rieke2015}) demonstrate the high signal-to-noise ratio and information-rich spectra obtainable for these objects \cite{Miles2023, Luhman2023, Beiler2023}. The aim of this work is to determine if the unprecedented quality of JWST observations allows for the precision abundance measurements required to disentangle the formation scenarios of stars, brown dwarfs, and planets. 

We present JWST NIRSpec observations with the G395H filter ($\sim$2.9 - 5.2 $\mu$m, R$\sim$2700) of the T8 dwarf 2MASS J04151954-0935066 (2MASS 0415-0935) obtained with Cycle 1 GO program 2124. This work represents the first precision abundance measurements from JWST NIRSpec observations at the maximum native resolution of R$\sim$2700.  We also explore the effect of combining the JWST full resolution spectrum with lower resolution spectra of the same object with the NASA Infrared Telescope Facility (IRTF) SpeX spectrograph \cite{Burgasser04} and the Infrared Spectrograph (IRS) of the Spitzer Space Telescope \cite{Suarez2022}.  We use the same retrieval framework successfully applied for low-resolution spectra of brown dwarfs \cite[e.g.,][]{Line2017, Zalesky2022} and recently modified for medium-resolution spectra \cite{Hood2023} (see Methods). 

\section{Results}\label{sec2}
\subsection{Agreement between retrieval model and G395H data}
\begin{figure*}[h]
         \centering
        \includegraphics[width=\linewidth]{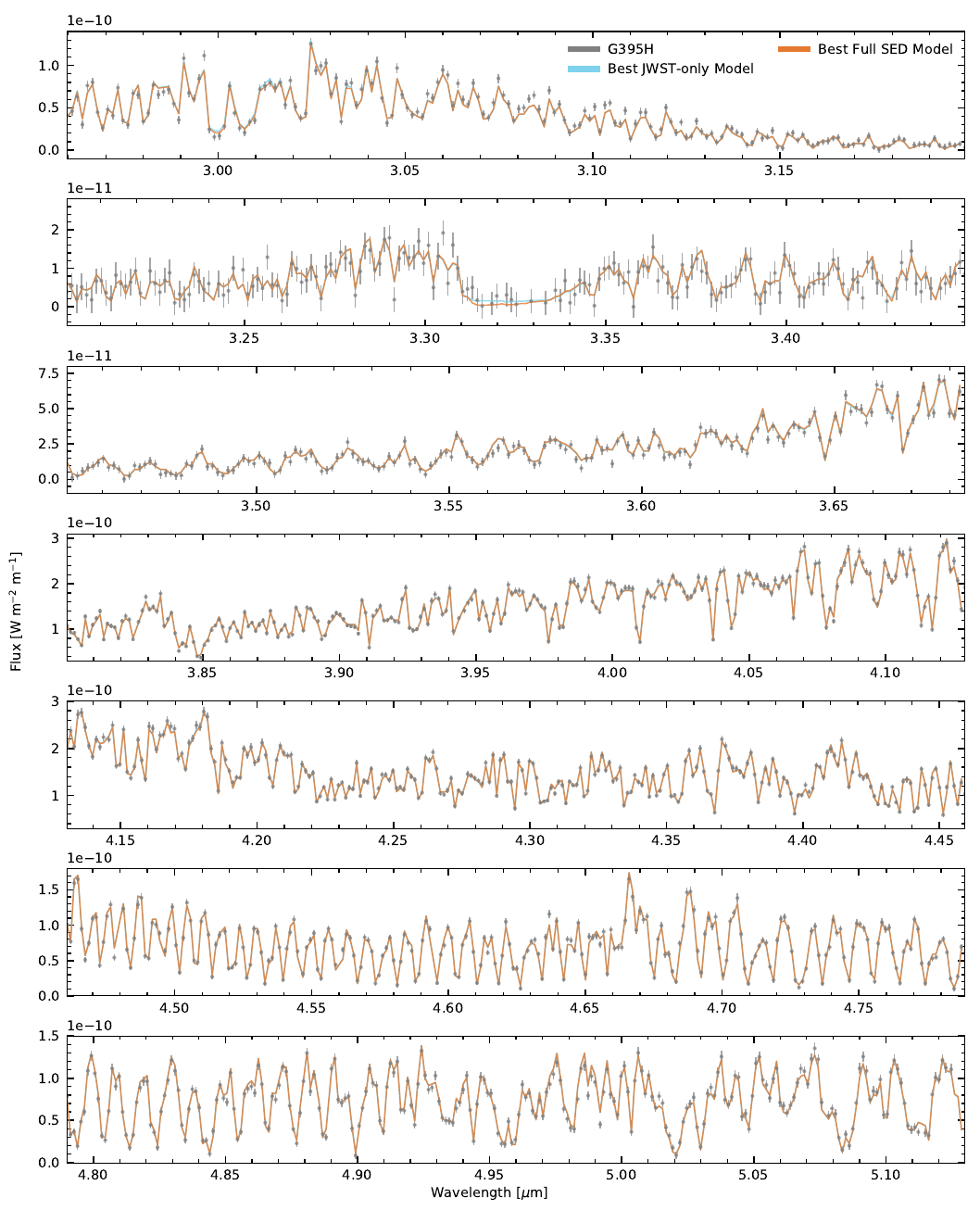}
  		\caption{\textbf{Observed NIRSpec/G395H spectrum of the T8 dwarf 2MASS 0415-0935 and the best-fit retrieval model.} The observed spectrum and associated error bars are shown in grey, the best-fitting model from a retrieval on just the JWST/NIRSpec data is shown in light blue, and the best-fitting model from a retrieval on all three datasets is shown in orange.}
  		\label{fig:ModelvData}
\end{figure*}

Figure \ref{fig:ModelvData} shows the observed NIRSpec/G395H spectrum compared to the best fit model spectra from our retrieval on solely the JWST data, as well as from the retrieval on all three data sets or the ``full SED" retrieval. Both models fit the observed spectrum well, and show only minor differences across these wavelengths. With the JWST data alone, we constrain the abundances of H$_{2}$O, CH$_{4}$, CO, CO$_{2}$, and NH$_{3}$ to within $\sim$ 0.15 dex, as shown in Figure \ref{fig:TPandposteriors}. We are also able to constrain the temperature--pressure ($TP$) profile, surface gravity, radius, radial velocity, and the ratio of $^{12}$CO/$^{13}$CO (discussed further below). The histograms indicate that H$_{2}$S abundance is not well-constrained, showing a long tail out to low abundances. We obtain only an upper limit on the PH$_{3}$ abundance, consistent with the lack of notable PH$_{3}$ spectral features. The NH$_{3}$ abundance constraint arises from a small feature at  $\sim$3 $\mu$m, previously tentatively identified \cite{Beiler2023} and confirmed in this work (see Methods for more information). 

\subsection{Effect of additional wavelength coverage}
Figure \ref{fig:TPandposteriors} shows the retrieved $TP$ profiles and posteriors of selected parameters for four retrievals on different combinations of data sets: JWST NIRSpec/G395H data alone, JWST and Spitzer/IRS, JWST and IRTF/SpeX, and the full SED retrieval. Adding in either the Spitzer or SpeX data yields more precise constraints that shift towards increase gravity and chemical abundances. However, including the Spitzer observations has little effect if the SpeX data are also included, although observations at these longer wavelengths with higher signal-to-noise might prove more useful. Adding in extended wavelength coverage also yields tighter $TP$ profile constraints in the deep atmosphere, particularly the SpeX data, reflective of the near-infrared observations probing the highest pressures and hottest temperatures. For the full SED retrieval, the $TP$ profile is constrained within $\pm$ 35 K for $\sim$0.4 - 40 bars. Furthermore, including the SpeX data in our analysis allows us to assess the alkali abundances, yielding a bounded constraint for K and an upper limit on log(Na) $\lesssim -4$. We also extrapolate the best fit model from the fits to each data set over the other wavelengths/instruments to test the ability of each data set to predict the others, as shown in Extended Data Fig. \ref{fig:NIRandFIR} and discussed further in the Methods section.

\subsection{Constraint on $^{12}$C$^{16}$O/$^{13}$C$^{16}$O}
\begin{figure}[h]
         \centering
        \includegraphics[width=0.75\linewidth]{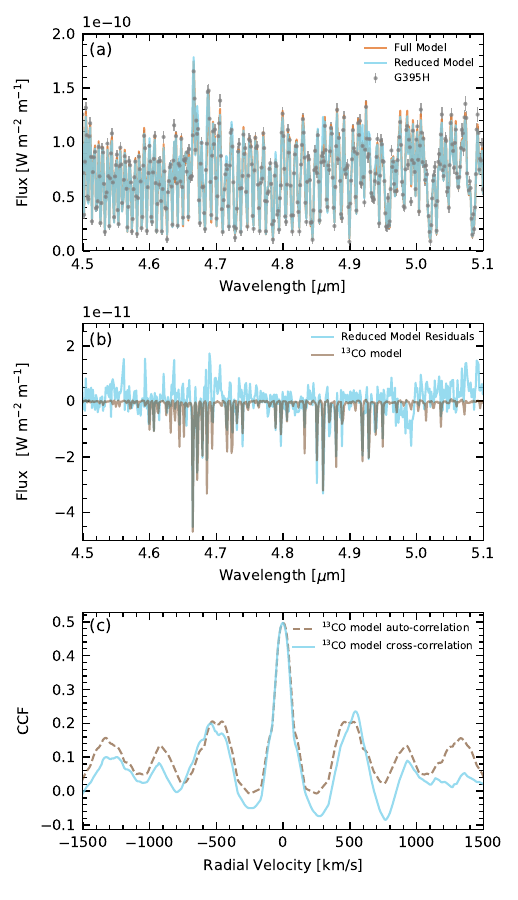}
  		\caption{\textbf{CO features in JWST/NIRSpec (G395H) data and $^{13}$CO cross-correlation detection.} (a) Portion of the G395H spectrum (grey data points) where CO is dominant, compared to the best fit model from a retrieval with both $^{12}$CO and $^{13}$CO in orange, and the best reduced model without $^{13}$CO in blue. (b) Residuals for the reduced model from (a), compared to a $^{13}$CO model (the difference between the best fit full model and the same model without $^{13}$CO) in brown. The reduced model residuals clearly overlap with the $^{13}$ CO absorption lines. (c) The cross-correlation function between the reduced model residuals and the $^{13}$CO model shown in blue. The auto-correlation function of the $^{13}$CO model scaled to the peak of the cross-correlation function is shown by the brown dashed line.}
  		\label{fig:Isotope}
\end{figure}

We include the $^{12}$CO/$^{13}$CO isotopologue ratio as a free parameter in each retrieval model. We are able to obtain a bounded constraint on this ratio with the NIRSpec/G395H data, finding $^{12}$CO/$^{13}$CO = 97.44$^{+8.78}_{-8.32}$ for the retrieval on the full SED of 2MASS 0415-0935. To validate this constraint, we follow the methods of \cite{Zhang2021} and perform a full retrieval including $^{13}$CO as well as a reduced retrieval without $^{13}$CO.  The best-fitting models of the full and reduced retrievals are shown in Figure \ref{fig:Isotope}a compared to the observed spectrum. We construct a $^{13}$CO model by taking the difference between the best fit model of the full retrieval and the same model without $^{13}$CO. The observational residuals of the reduced model are compared to this $^{13}$CO model in Figure \ref{fig:Isotope}b; the residuals clearly overlap the $^{13}$CO lines in multiple places. The cross-correlation function (CCF) between the residuals of the reduced model and the $^{13}$CO model is shown in Figure \ref{fig:Isotope}c, as well as the scaled auto-correlation function (ACF) of the $^{13}$CO model itself. The clear CCF peak at 0 km s$^{-1}$ indicates a strong detection of $^{13}$CO. We also calculated the Bayesian information criterion (BIC) of the best-fitting model for each retrieval to similarly determine if the inclusion of $^{13}$CO is justified. The difference in BIC between the two models, $\Delta$BIC = 357, indicates that including $^{13}$CO is strongly preferred ($\Delta$BIC $>$ 10 indicates very strong evidence against the model with higher BIC \cite{Kass1995}).

\section{Discussion}
\subsection{Physical Parameters of 2MASS 0415-0935}
From the best fitting spectrum and the directly retrieved parameters, we can calculate additional key diagnostic properties (see Methods), such as the bolometric luminosity L$_{Bol}$, effective temperature T$_{eff}$, mass (from the retrieved gravity and radius), metallicity [M/H], and the C/O ratio, as listed in Table \ref{tab:props}. Comparing our retrieved log(g) and calculated T$_{eff}$ to the Sonora Bobcat evolution models\cite{Marley2021} in Extended Data Fig. \ref{fig:EvolutionTracks}, we can see our constraints are consistent with an age of 4.0 - 6.0 Gyr and a mass of $\sim$ 37 - 45 M$_{Jup}$. 

The SED of 2MASS 0415-0935 has been studied by a number of authors, albeit with different data sets and techniques. Table \ref{tab:props} additionally lists the calculated physical parameters from two such studies. Both \cite{Fili2015} and \cite{Zalesky2022} find a lower effective temperature and larger radius than in the present work (though the error bars of \cite{Fili2015} are significantly larger), which balance out to yield similar L$_{bol}$. While our retrieved radius is smaller than previously reported for 2MASS 0415-0935, it is not so small as to be physically implausible, a common issue in brown dwarf retrieval studies \cite[e.g.][]{Lueber2022,Hood2023}. Though \cite{Zalesky2022} obtain a similar log(g) to our value, their solar metallicity and markedly supersolar C/O are inconsistent with the constraints presented here. As the retrieval framework used in this work is a modified version of that of \cite{Zalesky2022}, these discrepancies likely stem from the inclusion of longer wavelength data in addition to the SpeX spectrum used in that work, with which we can probe a wider number of chemical species in the atmosphere, along with opacity updates.

\subsection{Comparison to Grid of Radiative-Convective Equilibrium Forward Models}
To further contextualize our results, we compare the results of our retrieval to the Sonora Elf Owl grid of RCE forward models\cite{Mukherjee2023b} which self-consistently include the effects of disequilibrium chemistry. Chemical abundances in the visible atmosphere are governed by the interplay of chemical conversion and vertical mixing timescales - when the latter is shorter than the former, the species in question will be out of chemical equilibrium. For the Sonora Elf Owl models, the vertical mixing is parameterized with the vertical eddy diffusion coefficient K$_{zz}$, where a higher K$_{zz}$ will lead to more vigorous mixing and a shorter mixing timescale. The strength of this vertical mixing affects the abundances of many species considered in our retrieval, including H$_{2}$O, CH$_{4}$, CO, CO$_{2}$, NH$_{3}$, and PH$_{3}$. 

\begin{figure*}[h]
         \centering
        \includegraphics[width=\linewidth]{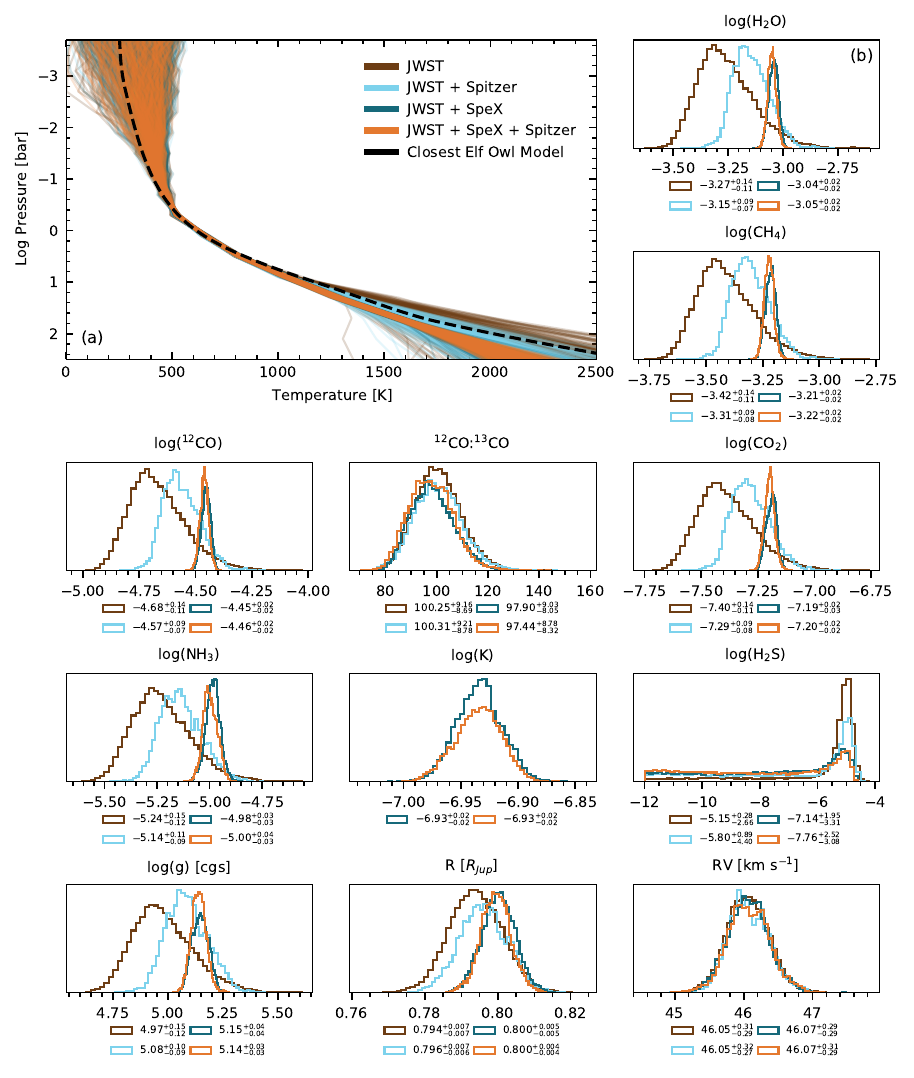}
  		\caption{\textbf{Retrieval results for different combinations of observations of 2MASS 0415-0935.} (a) The retrieved TP profiles. The results from retrieving on the JWST/NIRSpec data only are in brown, JWST/NIRSpec and the Spitzer/IRS are in light blue, JWST/NIRSpec and the IRTF/SpeX are in teal, and from including all three datasets are in orange. The black dashed line shows the TP profile from the Sonora Elf Owl grid for the closest grid point to our retrieved parameters. (b) The posterior distributions of selected parameters from each retrieval run, following the same colors as in panel (a). The legend lists the median value and 68\% confidence interval for each distribution. While the abundances of PH$_{3}$ and Na are also free parameters in our model, we only retrieve upper limits of log(PH$_{3}$) $\lesssim$ -7.5 for all cases and log(Na) $\lesssim$ -4 for the retrievals including the SpeX data.}
  		\label{fig:TPandposteriors}
\end{figure*}

We used a Bayesian grid fitting technique to compare all three datasets of 2MASS 0415-0935 to the Sonora Elf Owl grid (see Methods). Table \ref{tab:props} lists the resulting constraints for the grid parameters. While the effective temperature and radius from the Sonora Elf Owl fit are similar to those from our full SED retrieval, the grid model fit prefers significantly lower surface gravity, metallicity, and C/O. As shown in Extended Data Fig. \ref{fig:EvolutionTracks}, the decreased gravity at this effective temperature implies a much younger ($\sim$0.4 - 0.6 Gyr) and less massive ($\sim$ 15 M$_{Jup}$) object when compared to Sonora Bobcat evolution models. The $TP$ profile of the closest Elf Owl grid point is shown in Figure \ref{fig:TPandposteriors}a compared to our retrieved TP profiles. When the IRTF/SpeX spectrum is included, the retrieved $TP$ profiles deviate from the Elf Owl model deeper than $\sim$ 10 bars in a manner similar to \cite{Tremblin2019,Leggett2021}, potentially suggesting deviations from adiabatic convection.

\begin{figure*}[h]
         \centering
        \includegraphics[width=\linewidth]{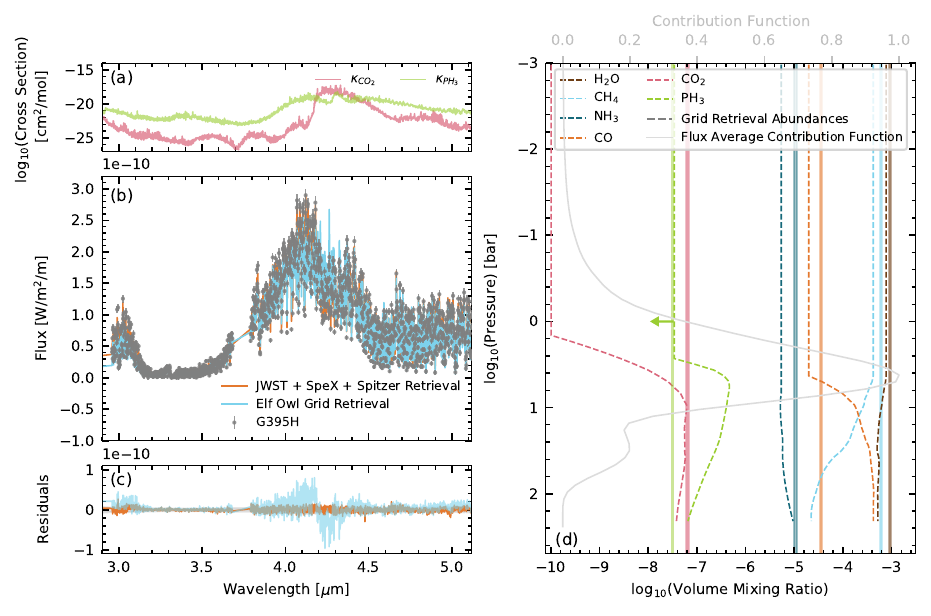}
  		\caption{\textbf{Comparison to grid retrieval results from the Sonora Elf Owl grid.} (a) The absorption cross-sections of CO$_{2}$ (pink) and PH$_{3}$ (green) at 1 bar and 650 K. (b) The JWST/G395H data in grey, compared to the best fitting model from our free retrieval on all three datasets in orange, and the best fitting interpolated spectrum from the Sonora Elf Owl grid in blue. (c) The observational residuals of both models from panel (b). The errorbars of the G395H data are represented by the grey shaded region. (d) The uniform-with-altitude mixing ratios of H$_{2}$O (brown), CH$_{4}$ (light blue), NH$_{3}$ (teal), CO (orange), and CO$_{2}$ (pink) from our retrieval on all three datasets are shown by the vertical shaded regions, which span the 68\% confidence interval of each abundance. The retrieved upper limit on the mixing ratio of PH$_{3}$ is indicated by the green line and accompanying arrow. The corresponding abundance profiles from the interpolated Sonora Elf Owl grid are shown by the dashed lines. Overplotted in light grey is the flux average contribution function.}
  		\label{fig:ElfOwl}
\end{figure*}

The best fit spectrum from the Elf Owl grid is shown in Figure \ref{fig:ElfOwl}b, compared to the NIRSpec/G395H data and the best model from our full SED retrieval. The model residuals from the data shown in Figure \ref{fig:ElfOwl}c highlight that the Elf Owl model particularly struggles from $\sim$3.8 - 4.3 $\mu$m, the region of the spectrum where CO$_{2}$ and PH$_{3}$ are important absorbers (representative cross sections of each molecule are shown in Figure $\ref{fig:ElfOwl}$a, references shown in Extended Data Table \ref{tab:opacity}). In particular, the model from the Elf Owl grid fit has too much PH$_{3}$ and too little CO$_{2}$ to match the data, as indicated by the residuals in Figure $\ref{fig:ElfOwl}$c. 

The chemical abundances as a function of pressure for the closest Elf Owl grid model to the fitted parameters are shown in Figure \ref{fig:ElfOwl}d, compared to the constant-with-height abundances from our full SED retrieval. Overplotted in gray is the flux average contribution function for our best fit model. For the regions of the atmosphere we are probing, our retrieved abundances for H$_{2}$O, CH$_{4}$, NH$_{3}$, and CO are all higher than those in the Elf Owl grid model. The largest discrepancies are CO$_{2}$ and PH$_{3}$, consistent with where the best Elf Owl model spectrum struggles to match the observed data. The difficulty in having enough CO$_{2}$ to match the observed features while reproducing the rest of the observed spectrum indicates the assumed chemical timescale for CO$_{2}$ in the Elf Owl model may be incorrect or the vertical mixing may be faster where CO$_{2}$ quenches in the atmosphere. Furthermore, 2MASS 0415-0935 joins a number of T and Y dwarfs for which PH$_{3}$ is expected but not detected \cite{Morley2018, Miles2020, Beiler2023} suggesting a need for major revisions to our understanding of phosphorous chemistry.

\subsection{Implications for Future NIRSpec/G395H Observations of Brown Dwarfs}
\textbf{Taking inventory of major carbon- and oxygen- bearing molecules:} Our retrieval results on solely the JWST NIRSpec/G395H observations of 2MASS 0415-0935 allow us to constrain all major carbon- and oxygen-bearing molecules (H$_{2}$O, CO, CO$_{2}$, and CH$_{4}$) to within $\sim$0.15 dex, allowing robust estimations of the object’s metallicity and C/O ratio (see Methods for more information). While the G395H observations alone can provide significant insight into the atmosphere of this brown dwarf, the posteriors for each parameter do  shift and tighten with the inclusion of data at other wavelengths. Adding in near infrared wavelength coverage in particular yields substantially better constraints on atmospheric parameters.

\textbf{Measuring $^{12}$CO/$^{13}$CO}: 2MASS 0415-0935 is the fourth and coldest substellar object with a measured $^{12}$CO/$^{13}$CO ratio. The measured $^{12}$CO/$^{13}$CO =  97$^{+25}_{-18}$ for the young L5 dwarf 2MASS J03552337+1133437 \cite{Zhang2021b} is similar to our measured value  ($\sim$97$^{+9}_{-8}$) for field-age 2MASS 0415-0935, both of which are close to the measured solar value of 93.5 $\pm$ 0.7 \cite{Lyons2018}. In contrast, both the widely-separated super Jupiter TYC 898-760-1 b and the hot Jupiter WASP 77 Ab show enhanced$^{13}$CO, with $^{12}$CO/$^{13}$CO = 31$^{+17}_{-10}$ and 20$^{+23}_{-10}$, respectively \cite{Zhang2021,Line2021}. Our ability to place tight constraints on the $^{12}$CO/$^{13}$CO ratio for 2MASS 0415-0935 indicates the strong promise of JWST NIRSpec/G395H measuring this isotopologue ratio in addition to molecular abundances for a large sample of substellar objects in order to better tease out different formation pathways.

\textbf{Testing our understanding of chemistry and vertical mixing:} The precise constraints on chemical abundances we retrieve for 2MASS 0415-0935 allow us to identify shortcomings with the Sonora Elf Owl grid of models in fitting the NIRSpec/G395H observations, particularly in matching observed PH$_{3}$ and CO$_{2}$ features. The increased spectral resolution of the observations presented here aids in disentangling the PH$_{3}$ and CO$_{2}$ features, which can blend and mask each other at lower resolutions\cite{Beiler2023}. Atmospheric retrieval studies on a large sample of cool brown dwarfs observed with NIRSpec/G395H will be an important step in identifying how our knowledge of the chemistry and physics that govern these atmospheres needs to be improved.

\clearpage

\begin{table}
\caption{Parameters of 2MASS 0415-0935 calculated from this work and previous studies.} \label{tab:props}
\begin{tabular*}{\textwidth}{lllll}
\hline
\hline
Parameter & Full SED   & Filippazzo et al.  & Zalesky et al. & Sonora Elf Owl  \\
 & Retrieval  & 2015 \cite{Fili2015} & 2022 \cite{Zalesky2022} & Grid Retrieval \\
\midrule
Wavelength Range ($\mu$m)& 0.95 - 20.5 & 0.76 - 7.59\footnotemark[1] & 0.95 - 2.5\footnotemark[2] & 0.95 - 20.5 \\
log(g) (cgs) & 5.14$^{+0.03}_{-0.03}$  & 4.83$^{+0.51}_{-0.51}$ & 5.10$^{+0.15}_{-0.25}$ & 4.51$^{+0.02}_{-0.01}$\\
Radius (R$_{Jup}$) & 0.800$^{+0.004}_{-0.004}$ & 0.95$^{+0.16}_{-0.16}$ & 0.94$^{+0.06}_{-0.06}$ & 0.80$^{+0.01}_{-0.01}$\\
Mass (M$_{Jup}$) & 36$^{+3}_{-3}$ & 33$^{+22}_{-22}$ & 47.8\footnotemark[3] & 8.4\footnotemark[3] \\
C/O & 0.53$^{+0.01}_{-0.01}$ & & 0.96$^{+0.09}_{-0.01}$ & 0.362$^{+0.002}_{-0.002}$\\
{[M/H]} (dex) &  0.28$^{+0.02}_{-0.02}$& & -0.02$^{+0.07}_{-0.10}$ & 0.05$^{+0.01}_{-0.01}$\\
log(L$_{Bol}$/L$_{Sun}$) & -5.70$^{+0.04}_{-0.01}$  & -5.74$^{+0.01}_{-0.01}$ & -5.75\footnotemark[4] & -5.72\footnotemark[4]\\
T$_{eff}$ (K) & 758$^{+18}_{-3}$ & 677$^{+56}_{-56}$ & 675$^{+9}_{-5}$ & 745$^{+3}_{-3}$ \\
\botrule
\end{tabular*}
\footnotetext[1]{Combination of optical and near-infrared spectroscopy with mid-infrared photometry (see \cite{Fili2015}).}
\footnotetext[2]{The IRTF/SpeX spectrum at R$\sim$100.}
\footnotetext[3]{Calculated from the reported radius and log(g).}
\footnotetext[4]{Calculated from the reported radius and T$_{eff}$.}
\end{table}

\section{Methods}\label{sec11}
\subsection{Observations and Data Reduction}
JWST cycle 1 GO Program 2124 (PI J. Faherty) obtained both NIRSpec G395H spectra and MIRI F1000W, F1280W, and F1800W photometry to fill out the peak of the spectral energy distribution (SED) and the tail of the SED for 12 brown dwarfs. NIRSPEC data was obtained using the F290LP filter, the G395H grating, the S200A1 aperture and the SUB2048 subarray.  The resultant wavelength coverage ranged from 2.87 to 5.14 micron with a resolution of $\sim$2700.  Acquisition images were first obtained for each target using the WATA method, the CLEAR filter, and the NRSRAPID readout pattern.  2MASS 0415-0935 was observed with NIRSPEC on 16 October 2022 with 11 groups per integration, 3 integrations per exposure and 3 total dithers for a summation of 9 total integrations in 168.488 seconds of exposure time.  Recorded time including overhead for the 2MASS 0415-0935 NIRSPEC observation was 1.03 hours.   

MIRI photometry was obtained with the F1000W, F1280W, and F1800W filters.  For each filter the FASTR1 readout pattern was chosen with a 2-point dither pattern.  2MASS 0415-0935 was observed with MIRI on 18 September 2022 using 5 groups per integration for F1000W, F1280W, and F1800W filters.  Total exposure time plus overhead for the MIRI observing of 2MASS 0415-0935 was 0.52 hours.  

For the reduction of all JWST data, we relied on the pipeline outputs which update with calibration data as it is received from the telescope.  

\subsection{Spectral Energy Distribution Construction} \label{sec:SED}
Key to the analysis contained within this work is working with a SED that can be input into the retrieval software. For this work we used the SED constructed by Alejandro et al. in prep.  In that work, the authors used the open-source package SEDkit (\cite{Filippazzo20}) along with the parallax from \cite{Dupuy12} and all JWST data plus literature spectra.  While the retrieval itself was only inclusive of the JWST G395H, SpeX, and Spitzer spectra, the full SED was constructed with photometry from MKO YJHK, WISE W1 W2 W3, JWST MIRI F100W, F1280W, F1800W, and Spitzer IRAC  as well as optical spectra from \cite{Burgasser03}, and Akari data from \cite{Sorahana12}. The full analysis of the SED for 2MASS0415-0935 will be discussed in Alejandro et al. in prep.

\subsection{Retrieval Model}
For our atmospheric retrieval analysis we used the CHIMERA framework which has been successfully applied previously to brown dwarf spectra \cite{Line2017, Zalesky2019}. We specifically use the version of CHIMERA\cite{Zalesky2022} which uses graphical processing units (GPUs), modified as outlined in \cite{Hood2023} for use on medium-resolution spectra. We use the affine-invariant Markov Chain Monte Carlo (MCMC) ensemble sampler package \textit{emcee}\cite{FM2013} to sample posterior probabilities. The free parameters in our model and our adopted prior ranges are listed in Extended Data Table \ref{tab:params}. We include uniform-with-altitude volume mixing ratios of H$_{2}$O, CH$_{4}$, CO, CO$_{2}$, H$_{2}$S, NH$_{3}$, and PH$_{3}$. For retrievals including the SpeX spectrum we also include mixing ratios of the alkalis K and Na, as they are expected to contribute important opacity at the near-infrared wavelengths. The opacity sources for each included chemical species are listed in Extended Data Table \ref{tab:opacity}. Extended Data Figure \ref{fig:Xsecs} shows the cross sections for each molecule at a representative pressure and temperature over the wavelengths of the G395H observations. As in \cite{Line2021}, we include the $^{12}$C$^{16}$O and $^{13}$C$^{16}$O lines separately weighted by the terrestrial ratio of $^{12}$C:$^{13}$C = 89:1 built into the HITRAN/HITEMP line lists. The CO isotopic abundance is parameterized as the log$_{10}$ of the isotopic ratio relative to terrestrial. 

We also include errorbar inflation exponents\cite{Line2015} for each data set analyzed, to account for underestimated data uncertainties or missing physics in the modeling framework.  There are 15 independent TP profile points, which are subject to a smoothing hyperparameter\cite{Line2015}. The TP profile points are interpolated onto a finer 70 layer pressure grid for the radiative transfer. Other free parameters include the surface gravity, a radius-to-distance scaling factor, and three cloud parameters: cloud volume mixing ratio, cloud pressure base, and sedimentation efficiency\cite{Ackerman2001}. For the cloud opacity we used Mie scattering theory assuming a Mg$_2$SiO$_4$ cloud with optical properties from \cite{Wakeford2015}. 

We generate the model spectrum piecewise at $\sim 10\times$ the spectral resolution of the incorporated datasets. For the G395H data, the spectral resolution of R$\sim$2700 is sufficient to potentially see effects from radial and rotational velocities. Thus, we incorporate these properties as free parameters as well, as described in \cite{Hood2023}. In brief, the radial velocity is applied using the \textit{dopplerShift} function from \textit{PyAstronomy}\cite{Czesla2019}.  The rotation velocity is applied using the \textit{fastRotBroad} function from \textit{PyAstronomy}, applied separately on the blue and red halves of the G395H model to minimize errors from not using the slower wavelength-dependent rotation kernel. The shifted and broadened model spectrum is then binned to the instrument wavelength arrays using a tophat kernel, and scaled to the observed flux using the (R/D)$^{2}$ parameter.

The choice of method used to smooth the forward model to the spectral resolution of the observations can affect the $v$ sin $i$ constraints from our retrieval. In addition to the tophat binning, we explored convolving the model with a Gaussian kernel meant to capture the wavelength-dependent resolving power of the instrument (assuming the instrument line shape is Gaussian and Nyquist sampled) and then interpolating the smoothed model onto the input data wavelength grid. With this method, we no longer constrain the $v$ sin $i$ of this object and the median radial velocity decreases by 3 km s$^{-1}$. However, no other posteriors significantly change.  As such, to avoid making as many assumptions about the exact instrument line profile, all reported values and posteriors are from retrievals using the tophat binning method. However, we caution that perhaps the resulting RV and particularly the $v$ sin $i$ constraints may not be reliable. Figure \ref{fig:CornerPlot} shows the corner plot summary of the posterior probability distribution, including the RV and $v$ sin $i$. 

\subsection{Sonora Elf Owl Grid Retrieval}
The Sonora Elf Owl grid includes cloud-free 1D RCE model atmospheres with vertical mixing induced disequilibrium chemistry across a large range of T$_{\rm eff}$ (275-2400 K), log(g) (3.25-5.5), the mixing parameter K$_{\rm zz}$ (10$^2$-10$^9$ cm$^2$/s), atmospheric metallicity (0.1$\times$ to 10$\times$ Solar), and C/O ratio (0.22 to 1.14). The model grid was computed using the open-sourced Python-based PICASO atmospheric model \cite{Mukherjee2023,Batalha19}. The Sonora Elf Owl grid is a successor to the Sonora Cholla grid \cite{karilidi2021} which simulated atmospheric chemical disequilibrium between T$_{\rm eff}$ of 500-1300 K by including variation in K$_{\rm zz}$ for solar composition atmospheres only. Other than increasing the T$_{\rm eff}$ coverage from 275-2400 K, the Sonora Elf Owl model has additionally varied atmospheric metallicity and C/O ratio across a large range of sub-solar to super-solar values. These variations are important because parameters like atmospheric metallicity and K$_{\rm zz}$ can have degenerate effects on the spectra of brown dwarfs.

For a more direct comparison with the retrieval results, the calculated TP profiles and atmospheric chemistry from the Elf Owl grid are first used to recompute the thermal spectra using the same gaseous opacities and radiative transfer code used for the retrieval analysis in this work. These recomputed high-resolution (R$\sim$ 30,000) spectra are then used to perform a Bayesian grid fitting analysis on the observed spectra of 2MASS 0415-0935. The Python Scipy based linear interpolating function "RegularGridInterpolator" \cite{scipy2020} is used to linearly interpolate the Elf Owl grid spectra at each wavelength point as a function of T$_{\rm eff}$, log(g), log$_{10}$K$_{\rm zz}$, metallicity, and C/O ratio. This spectral interpolator is then wrapped within the DYNESTY Bayesian sampler \cite{speagle20}. We assume uniform priors for T$_{\rm eff}$, log(g), log$_{10}$K$_{\rm zz}$, metallicity, and C/O ratio within the extent of these parameter values covered by the Elf Owl grid. We also use a uniform prior for the object radius between 0.6 to 1.2 $R_{\textrm{Jup}}$. Additionally, uniform priors between 0-50 km s$^{-1}$ are also used for both the rotational broadening velocity ($vsin(i)$) and radial velocity of the object. For each iteration of the Bayesian sampler, the interpolator is used to generate the model spectra for the drawn atmospheric parameters. The interpolated spectrum is then rotationally broadened using the \textit{fastRotBroad} module of the \textit{PyAstronomy} package \cite{Czesla2019}. The rotationally broadened spectra is also Doppler shifted with the \textit{dopplerShift} module of the \textit{PyAstronomy} package \cite{Czesla2019}. The high resolution spectra is then scaled by the sampled object radius and finally binned down to the wavelength bins of the observed data. The model spectra and the observed data are then used to calculate a log-likelihood metric which the sampler tries to minimise while it iterates to find the best-fit model. 

The obtained posterior distributions on the atmospheric and other parameters are then used to estimate the best-fit parameter values and their uncertainties. It should be noted that this technique does not take the model interpolation uncertainties into account while estimating the posterior distributions of each atmospheric parameter. Previous work has shown that the uncertainties on the estimated parameters can be somewhat boosted when the interpolation uncertainties are taken into account while fitting the data \cite[e.g.,][]{zjzhang21}.

\subsection{Calculation of Physical Parameters}
Table \ref{tab:props} lists a number of physical parameters of 2MASS 0415-0935 calculated from our retrieval on all three datasets (IRTF/SpeX, JWST/NIRSpec, and Spitzer/IRS), as well as the fit of these data to the Sonora Elf Owl grid of models and two previous studies of this object. To calculate these properties and their uncertainties from the parameters in our retrieval framework, we take 5000 random samples of the posterior. We calculate elemental abundances from our precise molecular volume mixing ratio constraints, with C/H = (CH$_{4}$ + CO + CO$_{2}$)/H, N/H = NH$_{3}$/H, O/H = (H$_{2}$O + CO + 2CO$_{2}$)/H, S/H = H$_{2}$S/H, and P/H = PH$_{3}$/H. We note that the total nitrogen abundance, and the resulting metallicity, is likely a lower limit as a significant amount of the object’s nitrogen could be in the form of N$_{2}$ (undetectable in the spectrum due to lack of absorption features) instead of NH$_{3}$.   As in previous works \cite{Line2021, Zalesky2022}, we account for potential depletion of atmospheric oxygen due to condensation by multiplying the oxygen abundance by 1.3 (the correction factor needed assuming 3.28 O atoms per Si atom from silicate cloud formation \cite{Burrows1999}).We do not include a correction factor when reporting the $^{12}$CO/$^{13}$CO ratio as the atmospheric carbon abundance should not be affected by condensation and the oxygen depletion should affect both isotopologues equally. However, this oxygen correction factor has been called into question by \cite{Calamari2022} and therefore better understanding of potential oxygen sinks in brown dwarf atmospheres may be needed to more confidently connect measured atmospheric oxygen abundances to the bulk value. 

From these elemental abundances, we calculate the metallicity [M/H] as the sum of the elemental abundances and the metallicity relative to solar [M/H] = log$_{10}$(M/H$_{0415-0935}$ / M/H$_{solar}$) assuming the solar abundances from \cite{Lodders2009} for consistency with the Sonora family of models. We also calculate the C/O ratio from the C and O abundances. To calculate the L$_{Bol}$ and corresponding effective temperature T$_{eff}$ implied by our retrieval results, we generate a low-resolution spectrum over 0.3 to 250 $\mu$m for each sample in our posterior. 

\subsection{Extended Results}

\subsubsection{3 $\mu$m NH$_{3}$ Feature}
Beiler et al.\cite{Beiler2023} recently reported a newly identified NH$_{3}$ feature at 3 $\mu$m in the JWST NIRSpec/PRISM spectrum (R$\sim$100) of the Y0 dwarf WISE J035934.06-540154.6, although the detection is tentative given the signal-to-noise and spectral resolution of the data as the NH$_{3}$ feature only impacts two data points. As shown in Extended Data Figure \ref{fig:Xsecs}, the \textit{Q} branch of the $v_{1}$ band of NH$_{3}$ overlaps with a window in CH$_{4}$ and H$_{2}$O opacity in this region. We can confidently confirm the presence of this 3 $\mu$m NH$_{3}$ feature in the NIRSpec/G395H spectrum of 2MASS 0415-0935. Extended Data Figure \ref{fig:NH3} shows the best fitting model from our JWST retrieval compared to the same model with significantly reduced NH$_{3}$ abundance, showing the importance of NH$_{3}$ opacity in reproducing the observed flux over $\sim$30 data points at this wavelength. This NH$_{3}$ feature also explains our ability to constrain the NH$_{3}$ abundance with the G395H spectrum alone as shown in Figure \ref{fig:TPandposteriors}. Adding in the Spitzer/IRS data which covers the strong NH$_{3}$ feature at 10.5 $\mu$m does slightly shift and tighten our abundance constraint, but not substantially. Thus, future NIRSpec/G395H observations hold promise to constrain the NH$_{3}$ abundance for a multitude of cool brown dwarfs even when longer wavelength observations are not available.

\subsubsection{Model Compared to Extended Wavelength Coverage Data}
For each retrieval, we produce the best fit model spectrum over the full set of observed wavelengths for comparison to the observed data for 2MASS 0415-0935, as shown in Extended Data Fig. \ref{fig:NIRandFIR}. For the JWST and JWST plus Spitzer retrievals, we assume the median alkali abundances from the full SED retrieval, as the modeled near infrared flux would be greatly overestimated without any alkali opacity. All four models are almost indistinguishable over the NIRSpec/G395H wavelengths, and only minor differences can be seen across the Spitzer/IRS region. The best fit models are most distinct in the near infrared, particularly in the peaks of \textit{y}, \textit{J}, and \textit{H} band. These discrepancies are consistent with the strong effect the SpeX data has on the retrieved constraints shown in Figure \ref{fig:TPandposteriors}. However, different assumed alkali abundances could bring the JWST and JWST plus Spitzer models into better agreement with the observations. 

\subsubsection{Elemental Abundances in Solar System Context}
We normalize the calculated elemental abundances of 2MASS 0415-0935 (see above) by the protosolar values for comparison with solar system values as in \cite{Atreya2016,Line2021}, with [X/H] = log$_{10}(\frac{n_{x}/n_{H}}{n_{x,\odot}/n_{H,\odot}})$. Figure \ref{fig:ElementalAbundances} shows these abundances compared to those of the solar system giants using values from \cite{Atreya2016,Li2020} and assuming the protosolar elemental abundances of \cite{Lodders2009}. The calculated nitrogen abundance from the retrieved NH$_{3}$ abundance is [N/H]=0.073$^{+0.007}_{-0.005}$; however, we plot this [N/H] as a lower limit since much of nitrogen content might instead be in N$_{2}$.  We obtain tight and bounded constraints for [C/H] = 1.36$^{+0.07}_{-0.06}$ and [O/H] = 1.17$^{+0.05}_{-0.05}$, consistent with a slightly super-solar metallicity and with a precision that in some cases greatly exceeds that for solar system objects.

\clearpage

\renewcommand{\tablename}{Extended Data Table}
\renewcommand\thetable{\arabic{table}}
\setcounter{table}{0}

\renewcommand{\figurename}{Extended Data Fig.}
\renewcommand\thefigure{\arabic{figure}}
\setcounter{figure}{0}

\begin{table}[h]
\centering
\caption{Free Parameters in Our Retrieval Model} \label{tab:params}
\begin{tabular}{lll}
\hline
\hline
Parameter & Description & Prior Range\\
\hline
log$_{10}$(f$_i$) & log$_{10}$ of the uniform-with-altitude volume & -12 - 0\\ 
& mixing ratios of H$_{2}$O, CH$_{4}$, CO, NH$_{3}$,H$_{2}$S, & \\
&    PH$_{3}$, Na, and K &  \\
$^{13}$CO$/$$^{12}$CO & log$_{10}$ isotopic ratio relative to terrestrial (1/89) & -3-3 \\
log(g) & log surface gravity [cm s$^{-2}$] & 0 - 6\\
(R/D)$^2$ & radius-to-distance scale [R$_{\rm Jup}$/pc] & 0 - 1\\
T(P) & temperature at 15 pressure levels [K] & 0 - 4000\\
$b_{j}$ & errorbar inflation exponent for each dataset & 0.01$\cdot$min(err$_{j}^{2}$) - 100$\cdot$max(err$_{j}^{2}$)\\
$\gamma$ & TP profile smoothing hyperparameter & 0 - $\infty$\\
log(Cloud VMR)   & log of the cloud volume mixing ratio & -15 - 0 \\ 
log(P$_{c}$)& log of the cloud base pressure & -2.8 - 2.3\\ 
f$_{sed}$  & sedimentation efficiency & 0 -10 \\
RV & radial velocity [km s$^{-1}$] & 0 - 100\\ 
$v$ sin $i$ & rotational velocity [km s$^{-1}$] & 0 - 100 \\
\hline
\end{tabular}
\end{table}

\begin{table}
\centering
\caption{Opacity Sources for Our Retrieval Model} \label{tab:opacity}
\begin{tabular}{ll}
\hline
\hline
Species & Opacity Sources \\
\hline
H2-H2, H2-He CIA & Richard et al. 2012 \cite{Richard2012}\\ 
H$_{2}$O & Polyansky et al. 2018 \cite{Polyansky2018} \\
CH$_{4}$ & Hargreaves et al. 2020 \cite{Hargreaves2020}\\
CO & Li et al. 2015 \cite{Li2015},$^{12}$CO and $^{13}$CO included separately \\
 & weighted by the built-in terrestrial ratio\\
NH$_{3}$ & Coles et al. 2019 \cite{Coles2019}\\
H$_{2}$S & Tennyson et al. 2012 \cite{Tennyson2012}, Azzam et al. 2015 \cite{Azzam2015}, isotopologues \cite{Rothman2013}\\
PH$_{3}$ & Sousa et al. 2015 \cite{Sousa2015} \\
K & Allard et al. 2016 \cite{Allard2016}\\
Na & Allard et al. 2019 \cite{Allard2019}\\
\hline
\end{tabular}
\end{table}

\begin{figure*}[h]
         \centering
        \includegraphics[width=0.72\linewidth]{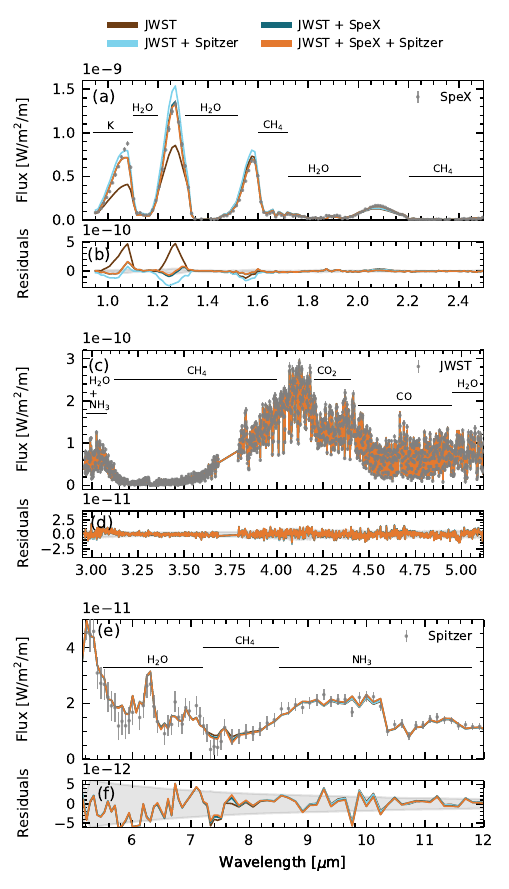}
  		\caption{\textbf{Comparison of median retrieval models to IRTF/SpeX, JWST/NIRSpec, and Spitzer/IRS observations of 2MASS 0415-0935.} (a), (c), (e) The models from each retrieval compared to the IRTF/SpeX, JWST/NIRSpec, and Spitzer/IRS observations, respectively. The median models from  retrieving on the JWST/G395H data only are in brown, JWST/G395H and the Spitzer/IRS are in light blue, JWST/G395H and the IRTF/SpeX are in teal, and from including all three datasets are in orange. While the Spitzer spectrum extends out to 20.5 $\mu$m, the models are indistinguishable past the wavelengths plotted here. (b), (d), (f) The model residuals from the IRTF/SpeX, JWST/NIRSpec, and Spitzer/IRS observations, respectively. The errorbars on the observed data are indicated by the shaded grey region.}
  		\label{fig:NIRandFIR}
\end{figure*}

\begin{figure}[h]
         \centering
        \includegraphics[width=\linewidth]{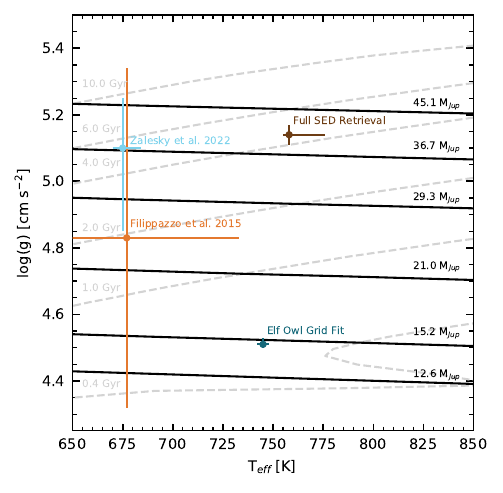}
  		\caption{\textbf{Comparison of retrieved physical parameters of 2MASS 0415-0935 to Sonora Bobcat evolutionary models\cite{Marley2021}}. Isochrones are shown in black and cooling tracks are shown in grey in the surface gravity-efferctive temperature plane. The median values and uncertainties of these properties from our full SED retrieval are shown in brown, while the results of the Sonora Elf Owl grid retrieval are shown in teal. Reported log(g) and T$_{eff}$ values from \cite{Fili2015} and \cite{Zalesky2022} are shown in orange and blue, respectively. The retrieved surface gravity and effective temperature agree with evolutionary models of age of 4 - 6 Gyr and a mass of 36.7-45.1 M$_{Jup}$, while the inferred mass from the retrieved surface gravity and radius is 36$^{+3}_{-3}$ M$_{Jup}$.  }
  		\label{fig:EvolutionTracks}
\end{figure}

\begin{figure}[h]
         \centering
        \includegraphics[width=\linewidth]{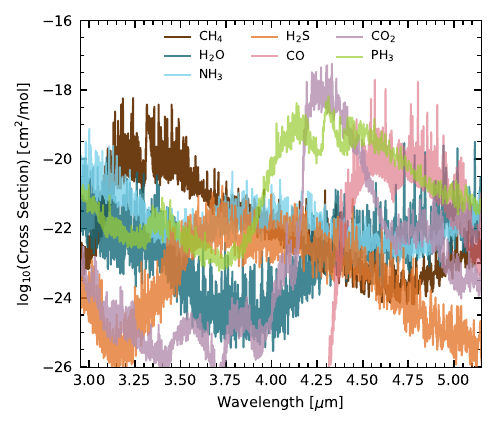}
  		\caption{\textbf{Absorption cross sections for molecular species included in our retrieval analysis over the wavelengths covered by the JWST/NIRSpec observations.} The plotted cross sections are for a temperature of 650 K and pressure of 1 bar.}
  		\label{fig:Xsecs}
\end{figure}

\begin{figure*}[h]
         \centering
        \includegraphics[width=\linewidth]{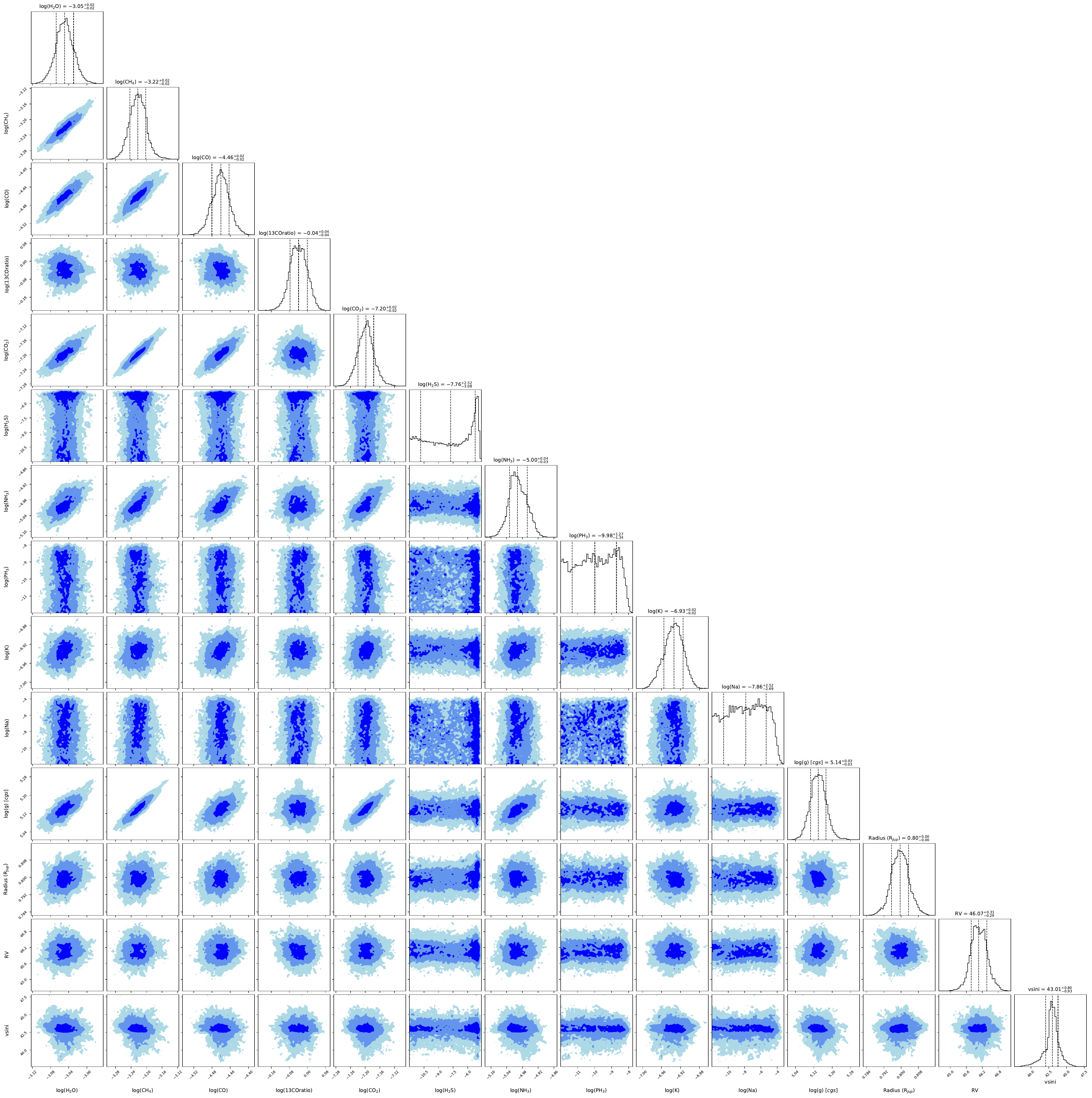}
  		\caption{\textbf{Summary of the posterior probability distribution of the retrieval analysis on all three datasets (JWST/NIRSpec, IRTF/SpeX, Spitzer/IRS).} For space constraints, we do not plot the posteriors for the temperature-pressure profile smoothing parameter $\gamma$, the error inflation terms for each dataset, or the cloud parameters, none of which showed strong correlations with other free parameters in our model. Here the CO isotopic abundance is parameterized as the log$_{10}$ of the isotopic $^{12}$CO:$^{13}$CO ratio relative to terrestrial.}
  		\label{fig:CornerPlot}
\end{figure*}

\begin{figure}[h]
         \centering
        \includegraphics[width=\linewidth]{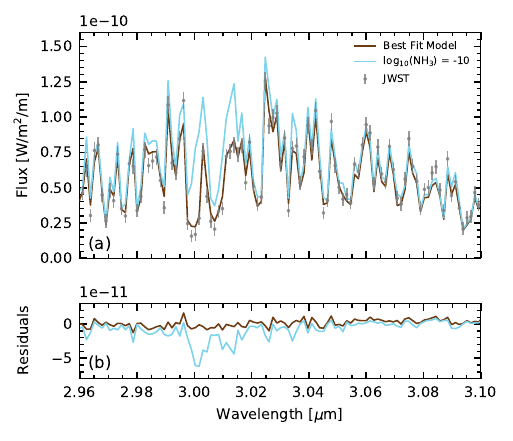}
  		\caption{\textbf{Evidence of NH$_{3}$ absorption around 3 $\mu$m.} (a) The JWST NIRSpec/G395H data (grey data points) compared to the best fit model from our JWST-only retrieval (brown) and the same model but where the volume mixing ratio of NH$_{3}$ has been reduced to log$_{10}$(NH$_{3}$) = -10 (blue). (b) The observational residuals of each model from panel (a). The reduced NH$_{3}$ model clearly struggles to fit the observed spectrum $\sim 3 \mu$m, confirming this small region as an NH$_{3}$ absorption feature.}
  		\label{fig:NH3}
\end{figure}

\begin{figure}[h]
         \centering
        \includegraphics[width=\linewidth]{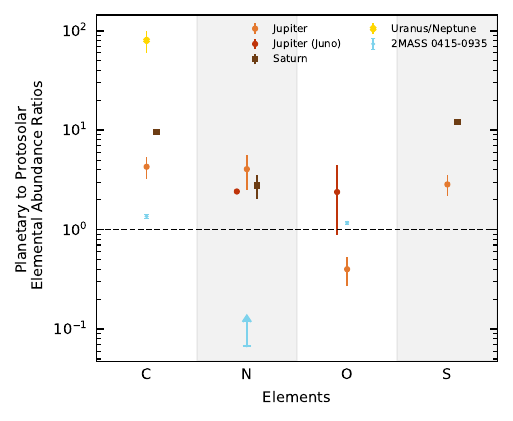}
  		\caption{\textbf{Comparison of elemental abundance constraints for 2MASS 0415-0935 to the solar system objects, adapted from \cite{Atreya2016, Line2021}.} Elemental abundances from \cite{Atreya2016} for Jupiter are shown in orange, Saturn in brown, and Uranus and Neptune in yellow, while more recent measurements for Jupiter from JUNO\cite{Li2020} are shown in red. The blue points and limits indicate the elemental constraints for 2MASS 0415-0935 derived from our retrieved abundances (see Methods). While we do retrieve tight constraints on the NH$_{3}$ abundance, the N/H value is shown as a lower limit as a significant amount of nitrogen may instead be in the form of N$_{2}$ in this object's atmosphere.}
  		\label{fig:ElementalAbundances}
\end{figure}

\clearpage

\backmatter

\bmhead{Acknowledgments}





\begin{appendices}


\end{appendices}


\bibliography{sn-bibliography}

\end{document}